# A 3D particle visualization system for temperature management

Lange B. [a], Rodriguez N. [a], Puech W. [a], Rey H. [b] and Vasques X. [b]
[a] LIRMM, 141 rue ADA, Montpellier, France;
[b] IBM, Rue de la vieille poste, Montpellier, France

**Abstract**

This paper deals with a 3D visualization technique proposed to analyze and manage energy efficiency from a data center. Data are extracted from sensors located in the IBM Green Data Center in Montpellier France. These sensors measure different information such as hygrometry, pressure and temperature. We want to visualize in real-time the large among of data produced by these sensors. A visualization engine has been designed, based on particles system and a client server paradigm. In order to solve performance problems, a Level Of Detail solution has been developed. These methods are based on the earlier work introduced by J. Clark in $1976$[1]. In this paper we introduce a particle method used for this work and subsequently we explain different simplification methods applied to improve our solution.

Keywords: 3D Visualization, Sensors, Particles, Client/Server, Level Of Details

## 1. INTRODUCTION

In this paper, we present a method to produce a 3D visualization for analyzing and managing temperature. Data are extracted from sensors located in the IBM Green Data Center in Montpellier, which provides many different types of information like temperature, pressure or hygrometry. In our system, sensors are placed in a virtual room and the internal space is modeled using particles. The main constraint here is to produce a real-time rendering. However, latency appears du to the number of vertices. In this paper, we use a solution called LOD (Level Of Detail) to produce multi resolution 3D objects. This solution has been introduced in 1976 by J. Clark [1] . In this paper, J. Clark introduces the use of several mesh resolutions to simplify the 3D scene complexity. In our work, we use various simplification methods to provide interactive rendering and allows rendering the most important part of data extracted from sensors. In this paper, we describe how we create a room, and the methods used to produce different resolution visualization.
In Section 2, we introduce related work on particles systems and LOD. In Section 3, we expose our solution to simplify particles system. In Section 4 we give some results and finally, in Section 5 we present our conclusions and future work.

## 2. RELATED WORK

In this section we present several previous works concerning data visualization, particle systems and level of detail methods.

Some previous work present solutions to visualize large data flow extracted from mantle convection. M. Damon et *al*. [2] and K. E. Jordan et al. [3] present interactive viewers for this kind of data. These data are computed by using Hight Performance Computing (HPC) and visualized on a large display. The rendering is calculated by using another HPC. The data flow is very important and a real-time 3D simulation is hard to obtain. W. Kapfer and

T. Riser [6] introduce how to use particle system to visualize astronomic simulation, particles representing space objects. The number of particles is extremely important for computing motion in real-time. GPU computing is preferred to render instead of a common HPC solution. To display their data, they have developed their own 3D graphical engine. The space objects are represented by point sprite instead of sphere. Lights are used to give a spherical aspect to the point sprite. This solution allows to render more stars than spherical object method. The 3D engine provides different rendering methods to group space objects: cell simplification or extraction of isosurface. The use of GPU seems quite well for a particle solution, parallel processing allows to render large data; the astrological data seems to be well suited.

In 1976, J. Clark introduces Level Of Detail (LOD) concept [1]. LOD consists to produce several resolution meshes for using them at different distance from the camera. Firstly, designer produces these meshes. First algorithms, in 1992 Schroeder et al. developed a method by decimation for simplify the mesh [7]. It analyses mesh geometry and evaluates the complexity of triangles. Vertices are removed if only constraints set by the user are respected. Vertices are removed and gaps are filled using triangulation. These algorithms of simplification are not enough to simplify mesh efficiently because shape is not always totally respected. D. Luebke, in 1997, has proposed a taxonomy of mesh simplification [8]. He presented the most used algorithms. He extracted different ways to use each algorithm. But in this paper, only one solution works with volumetric mesh [9]. T. He et al. propose a method based on voxel simplification by using a grid for clustering voxels. A marching cube [10] algorithm was applied to produce a surface mesh. But this simplification algorithm did not preserve the shape of the mesh. In our work, we look for point cloud simplification. Indeed, previous methods which deal with simplification for surface point cloud like [11-13] are not adapted to our case. All of these methods produce LOD for surface mesh and point cloud is extracted from scanner.

## 3. PROPOSED APPROACH

This section presents the different methods that are used to visualize a kind of data from Green Data Center (GDC). The main goal is to be able to visualize in real-time the evolution of temperature in the data center. For this, we use a special particle method. Particles are located using a segmentation algorithm based on Voronoï cell extraction and Delaunay triangulation. The latency due to the large flow of particles is avoided by using a client server paradigm. We improve our solution by using LOD methods to simplify rendering.

## 3.1 Particle systems

Rooms are the bases of our study. For modeling a room, we extract the shape of the space representation which is composed by a box with three measures: length ($l \in \mathbb{R}$), width ($w \in \mathbb{R}$), height ($h \in \mathbb{R}$). Sensors are represented by $S = \{S_1, \ldots, S_M\}$, where $M$ is the number of sensors. Sensors $S_i (i \in \{1, \ldots, M\})$ are placed on the space on a layer $L \in \mathbb{N}$ and have a location represented by: $\{X_i, Y_i, L_j\}$ with $X_i \in \mathbb{R}, Y_i \in \mathbb{R}$ and $j$ is the layer used. For modeling the space inside a room, we use a particle system instead of 2 D map representations which have some lacks.[14] Actually 2D map does not allow having a real visualization of space. A particle visualization gives a better efficiency for modeling space. We use a large number of particle to represent the entire space. $N \in \mathbb{N}$ represents the number of particles in the room. It can be calculated using:

$$N = \frac{((l+1) \times (h+1) \times (w+1))}{\delta^3} \quad (1)$$

where $\delta \in \mathbb{R}$ is the space between particles. The particle grid is regular. In this model, three layers of temperature sensors compose rooms. They are defined according to their real locations in the data center. Figure ?? presents the different layers of sensors in the data center.

Particles carry information, and flow motion can be simulated if needed by changing the value of particles and the computational cost is inferior.

## 3.2 Segmentation algorithms

In our solution, each sensors has an influence on surrounding particules. To calculate the set of particles in the sensor range, we use two methods: Voronoï cells extraction and Delaunay triangulation.

Voronoï cells is a method to extract a partition of space[15]. This method is available for $\phi$ dimensions where $\phi \in [1, +\infty]$, but most of implementations are done in 2D. Tools for extracting 3D Voronoï diagrams exist: Voro++[16] and QHull[17] but particles are discrete and these solutions are not suitable because they extract Voronoï diagram in a continuous way. Then we designed our own method based on sphere expansion. We search nearest sensors for each particle. This part allows to weight particles outside the sensors mesh. A second method to weight the interior of the sensors mesh is used. We extract the mesh tetrahedron of sensors using the Delaunay triangulation implemented in QHull[17]. This method was used to analyze the location of particle. We compute the exact location using ray tracing[18] on the soup of tetrahedron. First, we search the nearest particles inside the hull of each tetrahedron. We extract the normal of each face of tetrahedron and we apply these normals on each particle. If the ray cuts three faces or more, the particle is inside the tetrahedron. This method is cost expensive and done in preprocessing. Moreover, particles are static and position didn't need to be update.

## 3.3 Client server paradigm

To improve computation, a client server paradigm is used. We define a low cost communication protocol to transfer data from a server to a client. Server computes the modification of particles and the client displays the results. This protocol works in five steps. These steps are: sending header, sending sensor data, sending particle data, sending footer and receiving acknowledgment/command from client. At each step, the server waits the acknowledgment from the client. We develop two ways to send data. The first sends the entire point cloud (sensors and particles). The biggest problem of this method is the transmission of data. Sensors are sent with their coordinates and their value. We encode these data in bit words. For the particles data, the same method was used. The footer was sent for closing the communication. The second method is used to reduce efficiently the communication cost. We only send modified sensors and particles. The id and the new value is sent instead of coordinates. The last step is the command sent by the client. It allows the user to interact with the server. We use it to modify the camera viewpoint.

### 3.4 Level of detail for particles

Level of detail (LOD) is one of the most important methods in computer graphics. It allows to solve rendering problems or performance problems. This method consists by producing several resolution of a 3D object. In our works, we use some features to define the object resolution: hardware and viewpoint. Hardware and viewpoint do not need the same data structure and we need to recompute it for each modification of the viewpoint or when hardware changes. LOD was defined by two problems statement. The first one uses a sample of original points, the second one uses a new point data set. In this part, we define six methods to produce LOD. The four first methods are for the client, the other are for the server.

Problems statement:
For this two approaches, we have a set $\omega$ of Vertices $V$, $V = \{V_1, \ldots, V_\omega\}$. Each vertex is defined in $\mathbb{R}^3$. Simplify a mesh using a sample vertex means $\omega > \omega 2$, where $\omega 2$ is the size of the second data set. For approach 1, we obtain a new object $V2 = \{V2_1, \ldots, V2_\omega\}$ with fewer point than V but V 2 is a subset of V. For approach 2, we obtain a new object $V3 = \{V3_1, \ldots, V3_\omega\}$ with fewer point than V but each point in V 3 is a new vertex.

In Section 2 we have presented methods to produce simplification. A few were designed for volumetric simplification. In this section, we propose several methods to produce different volumetric simplifications on our client. We develop four approaches to simplify 3D objects: clustering, neighbor simplification and two approaches based on server. Clustering method was based on He et al. [9] works, it consists of clustering particles using a 3D grid. Cells sizes of grid are set depending to the viewpoint of the camera. Clusters were being weight with the average of the different values of particles. The position is the barycenter of these particles. Figures 1(a)-1(e) give some examples of simplification using clustering solution. Figure 1(a) present the original point of cloud mesh. Figure

1(b) and 1(d) give two different methods for clustering. And finally, Figure 1(c) and 1(e) give the results of clustering methods.

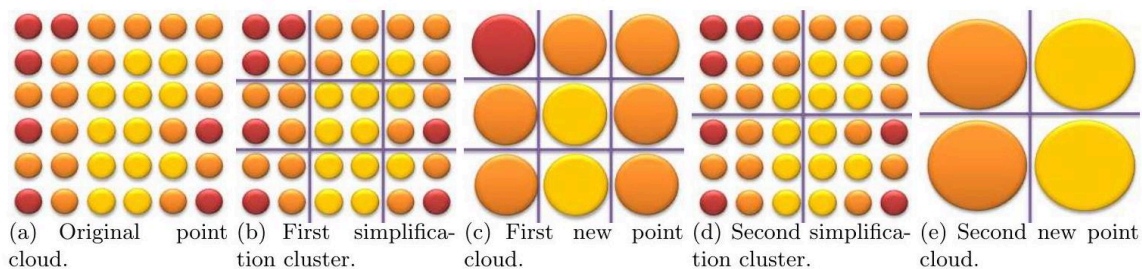

(a) Original point cloud. (b) First simplification cluster. (c) First new point cloud. (d) Second simplification cluster. (e) Second new point cloud.

Figure 1. Clustering method for simplification point cloud.
The second solution used is based on neighborhood extraction. Before runtime, we extract all neighbors of a particle. We measure the distance between each particle. Some optimization can help to decrease complexity: we can estimate easily in our structure which particle is closer to another one (using the fact that particle grid is regular). After this, we extract the main value of particles. We explore each neighbor of particles and we keep the most important. In some cases, the most important can be the high values, in other the low values and in other both of them. This solution is able to produce a low resolution model with the most important information structure. Several low resolution models are created by exploring deeper in neighborhood. Figures 2(a)-2(c) illustrate a neighbor, and two simplifications of this mesh.

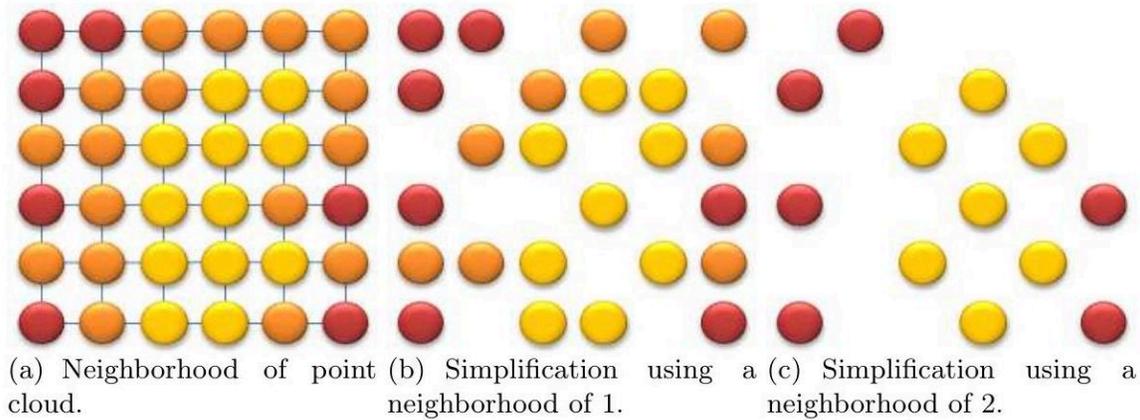

(a) Neighborhood of point cloud. (b) Simplification using a neighborhood of 1. (c) Simplification using a neighborhood of 2.

Figure 2. Neighbor method for simplification.

Other methods were based on server instead of client. Client sent via TCP connection his viewpoint. The server recomputes the particles structure and recreates the entire structure. With this solution, it is possible to produce a point cloud resolution depending on hardware. Figure 3(a) presents particles rendering with a distance of 2 from the camera. Figure 3(b) is the decimation produced with a distance of 3 and Figure 3(c) is a distance of 1.

Another method was based on Voronoï diffusion of temperature. The bandwidth for transmitting data is limited. We developed Voronoï temperature diffusion to solve this communication. In this approach, we update data using sphere expansion. Each time, we update particles depending on their distance from sensors. The more particles are distant from sensors the later they will be refreshed. This method sends only modified particles. The bandwidth is saved and the visualization gives a flow effect. Figure 4(a) represents values at time 0. At time 1, values of sensors change, 4(b). After time 2, we update a first range of particles 4(c) and finally the second range 4(d).

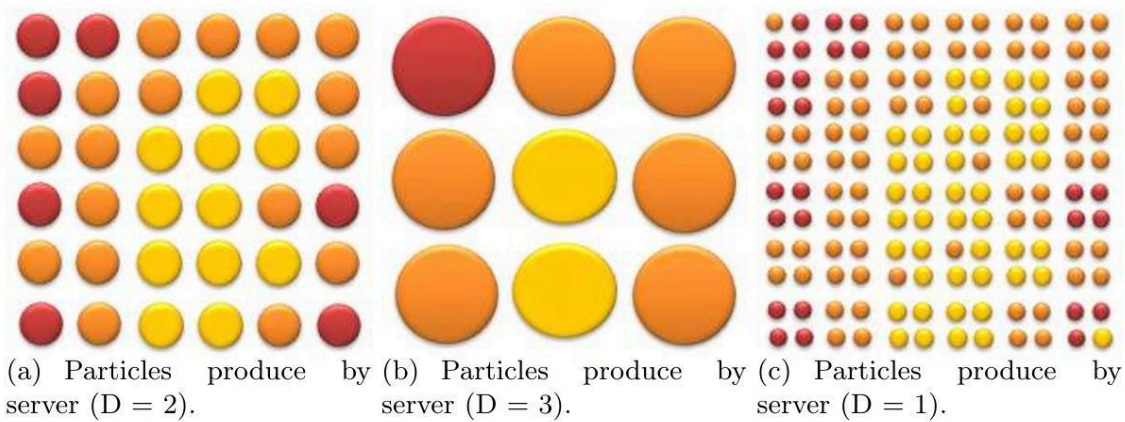

(a) Particles produce by server (D = 2). (b) Particles produce by server (D = 3). (c) Particles produce by server (D = 1).

Figure 3. Particle simplification using server and distance.

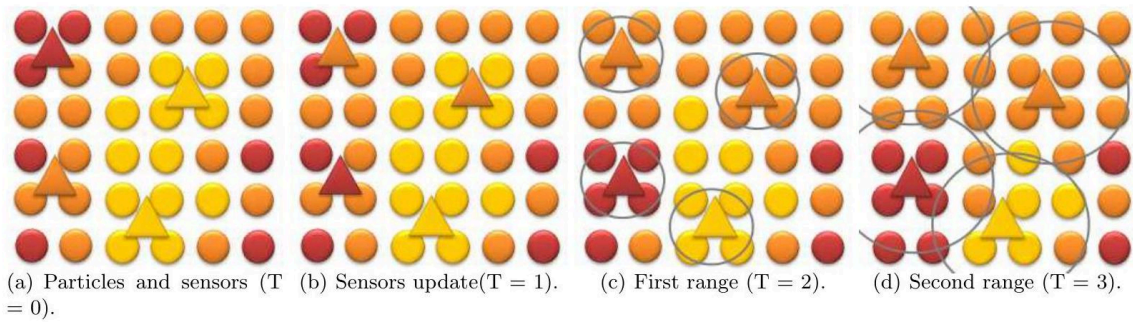

(a) Particles and sensors (T = 0).  (b) Sensors update (T = 1).  (c) First range (T = 2).  (d) Second range (T = 3).

Figure 4. Simplification using bandwidth size.

# 4. EXPERIMENTAL RESULTS

The data are extracted from two rooms of the IBM data center. Firstly, we present our method for rendering the room, and later we present our results using Level Of Detail methods.

## 4.1 Data visualization

We want to visualize and manage the consumption of a data center. For the visualization, we want to use an IFC viewer. But the IFC model for GDC is not available yet. Data center extraction of the room space is for the moment done by hand. The room is empty and was represent by a simple shape a box with 4 meters length, 3 meters width and 2.5 meters height. We use point cloud visualization based on particle paradigm. We use the two rooms of the data center and we put the same number of particles (30000) and 35 sensors distributed on three layers at 1 meter, 2 meter and on the ground. We define high and low temperature regarding the real sensors value. Figure 5(a) presents temperature color scale, Figure 5(b) and Figure 5(c) present data center sensors.

The next step is to interpolate data from sensors. For this, we extract the sensor mesh. We use QHULL to produce a soup of tetrahedrons. Particles need to be located. We can determine which tetrahedron is the nearest, we extract the box hull of tetrahedron and we apply for each particle the norms of each tetrahedron face. If these rays cut three or more faces, then particle is inside the tetrahedron. With this method, we can determine exactly the location of each particles regarding to the tetrahedrons, a weight is given to them easily. It was used to apply a coefficient to the value of each vertex of tetrahedron. For the outside particles, another solution was used: Voronoï cells. This method is based on a discrete extraction of Voronoï cells. We use our own method because other method like Voro ++ or QHull extract Voronoï diagram in a continuous way.

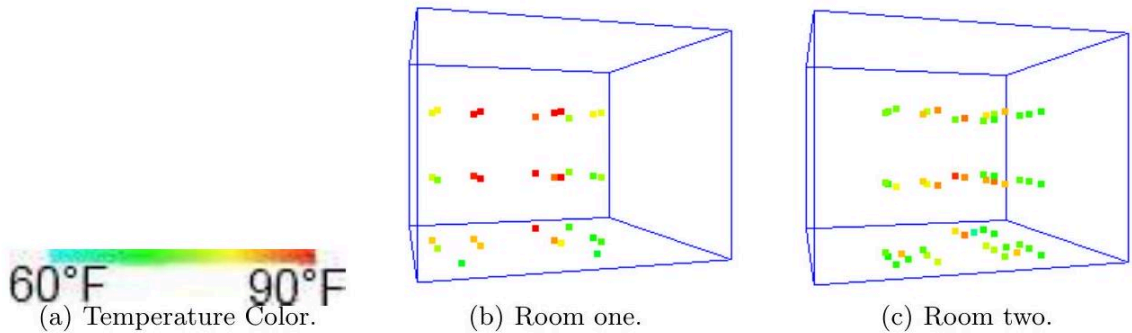

(a) Temperature Color.   (b) Room one.   (c) Room two.

Figure 5. Data use to model the system.

## 4.2 Level of details

In the earlier days of this project, first solution proposed gives a low frame rates, about 15 FPS (Frame Per Second): visualization was not in real-time (real-time is about 24 FPS ). For solving this problem, we define a client server paradigm. This solution allows to produce a real-time rendering on the client. Figure ?? gives an example of LOD for particles. We use Openscenegraph [20] as a 3D engine. It owns several features useful in LOD. A special object is defined to manage multi-resolution model. It calculates the distance of the object from the camera. For our experimentation we use five resolutions of mesh. The first mesh was the original mesh, it is set at 0 to 500 . The next mesh was set at 500 to 1000 , the next at 1000 to 1500 and the other at 1500 to 2000 . These three meshes were constructed by specific LOD methods: clustering and significant vertices. Clustering defines a 3D grid inside the room. The size of each cell depends on the viewpoint location. The size of the cluster depends on the visibility of the clustered particles. First results are given Figure 6(a) and 6(b). Value of cluster is an average of clustered value. The number of points of the final mesh depends on the grid size. Table 1 shows the results at several distances.

|       | $D = 0$ to 500 | $D = 500$ to 1000 | $D = 1000$ to 1500 | $D = 1500$ to 2000 |
|-------|----------------|-------------------|--------------------|--------------------|
| $C = X$ | 30000 | 3900 | 240 | 36 |

Table 1. Results of clustering simplification.

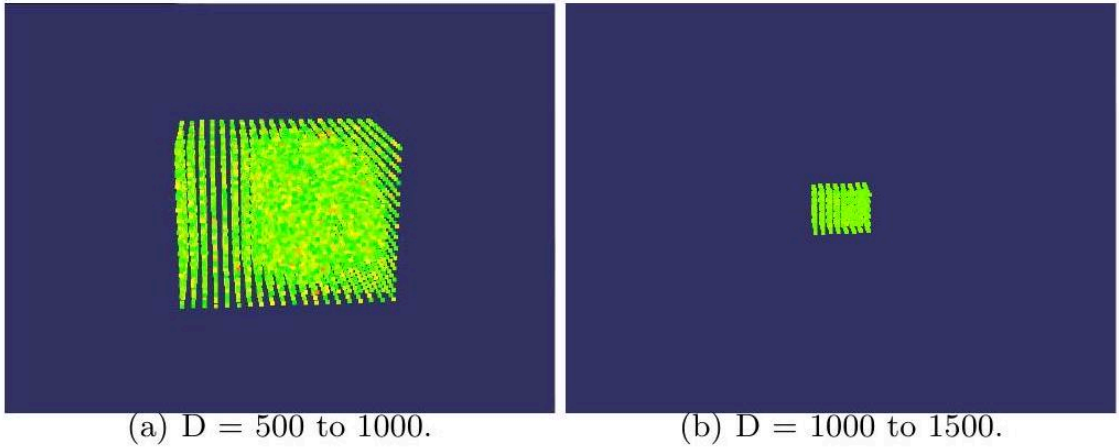

(a) $D = 500$ to $1000$.      (b) $D = 1000$ to $1500$.

Figure 6. Clustering visualization algorithms.
Significant points method extracts the neighbors for each particle. We extract the highest and lowest temperatures, by exploring the neighborhood of a particle, in order to have significant vertices of the model. For the first step of simplified model we explore neighbor. For the second model, we explore neighbor and neighbor of neighbor, etc. This solution simplifies drastically the model. First results are given Figure ??-??. Table 2 shows the number of vertices at several distance.

|  | D = 0 to 500 | D = 500 to 1000 | D = 1000 to 1500 | D = 1500 to 2000 |
|---|---|---|---|---|
| C = X | 30000 | 22950 | 4554 | 3524 |

Table 2. Results of neighbor simplification.

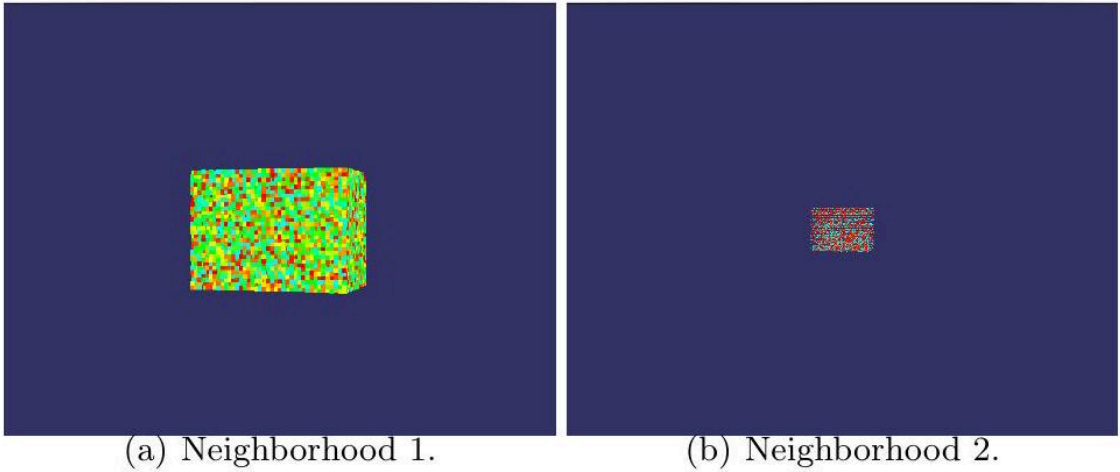

(a) Neighborhood 1.    (b) Neighborhood 2.

Figure 7. Clustering visualization algorithms using neighbor.

The first server solution receives orders from client as presented Section 3.4. We calculate the viewpoint distance and we send data according to it. A new structure is recalculated if the camera is too far from the object. After the recomputing, we send the new data. This solution allows the user to receive more or less data according to its distance to the object. Table 3 shows some different resolutions produced with this method.

|  | D = 0 to 500 | D = 500 to 1000 | D = 1000 to 1500 | D = 1500 to 2000 |
|---|---|---|---|---|
| C = X | 120000 | 30000 | 7500 | 1875 |

Table 3. Several resolution of model.

Another solution is to use bandwidth latency. We send data at several times, we do not send the entire set of data but only modified particles. We send at first time the sensors data, and subsequently we send a range of data (the nearest). After few minutes, all data are sent. This solution gives good results, and simulates a thermal diffusion in the whole structure of particles. Figure 8(a)-8(c) illustrate this method.

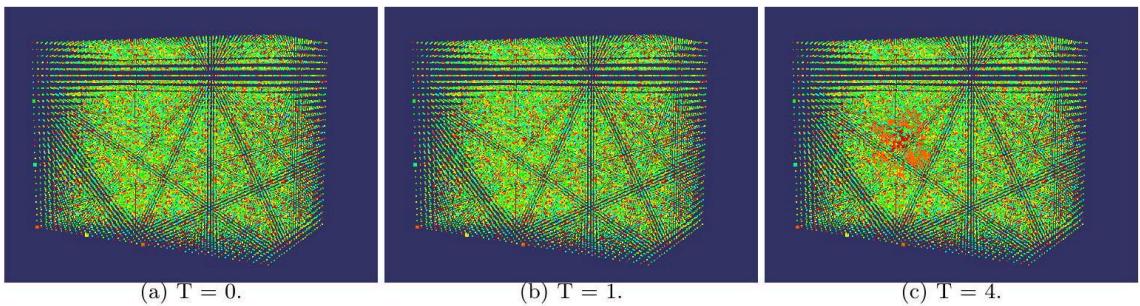

(a) T = 0.    (b) T = 1.    (c) T = 4.

Figure 8. Bandwidth simplification.

## 5. CONCLUSION

In this paper, we have presented a method to visualize sensors data extracted from a Green Data Center. This approach produces interpolation visualization for managing and visualizing data. This interpolation used a Delaunay triangulation and a cell extraction based on Voronoï . An unusual way of use particles helps to process data. First results present the solution proposed to visualize the inside of a GDC space. The second results proposed in this paper aim to improve the rendering.
For this, first step introduces a client/server protocol a second step illustrates methods to simplify the model. With these different approaches we improve the rendering time, preserving most important data are kept. In future works, we will work on data "dressing". We want to find a way to improve rendering of the scene using meatballs or marching cube algorithms. A main constraint of this work is real-time computation. Future work also concern to add rooms to the visualization. At present, we only visualize a single room. We want to visualize building, and complex form, by using an IFC loader.

## ACKNOWLEDGMENTS

We want to thanks the PSSC (Products and Solutions Support Center) team of IBM Montpellier for having provided the necessary equipment and data need for this experimentation. And we thank the FUI (Fonds Unique Interministriel) for their financial support.

Further author information:
Lange B.: E-mail: benoit.lange@lirmm.fr
Rodriguez N.: E-mail: nancy.rodriguez@lirmm.fr
Puech W.: E-mail: william.puech@lirmm.fr
Rey H.: E-mail:REYHERVE@fr.ibm.com
Vasques X.: E-mail: xavier.vasques@fr.ibm.com